\documentclass{article}[10pt]
\usepackage{abstract}
 \pdfoutput=1
\usepackage[latin1]{inputenc}
\usepackage[T1]{fontenc}
\usepackage{graphicx}
\usepackage{amsmath}
\usepackage{amsfonts}
\usepackage{amssymb}
\usepackage{amsfonts}
\usepackage{bm}
\usepackage{amsthm}
\usepackage{subfig}
\usepackage{tabularx}
\usepackage{amsbsy}
\usepackage{authblk}
\usepackage{lipsum} 
\usepackage{fancyhdr}
\usepackage{booktabs}
\usepackage{caption}
\usepackage{tabularx}
\usepackage{multicol}
\usepackage[a4paper,top=2cm,bottom=2cm,left=3cm,right=3cm]{geometry}
\begin{document}
\begin{flushright}
 Draft of:21 July 2016
\end{flushright}
\begin{center}
 {\bf Analysis of strategic and deterrence equilibrium by modeling with a Van der Waals gas.}\\
 Fabrizio Angaroni\footnote{Email: f.angaroni@uninsubria.it}
\\
 Center for Nonlinear and Complex Systems,
Universit\`a degli Studi dell'Insubria, via Valleggio 11, 22100 Como, Italy\\
Istituto Nazionale di Fisica Nucleare, Sezione di Milano,
via Celoria 16, 20133 Milano, Italy\\
Maurizio Martellini\footnote{Email: maurizio.martellini@uninsubria.it}
\\
Insubria center on international security (I.C.I.S.), via Natta 14, 22100 Como, Italy\\
\end{center}
\begin{flushleft}
 For PhD. program of Universit\`a degli Studi dell'Insubria
\end{flushleft}

\section{Abstract}
Physics is about the mathematical previsions of natural events.\\
Physicists are very careful to make predictions about the actions of living beings, as
living beings have a non-trivial behaviour and it is very hard to describe it with equations.\\
Anyway in modern literature some work of this kind emerges. Most of this work studies  the  behaviour of simple microorganisms or they are based on
a large number of collected data in order to make probabilistic or statistical studies.\\
Neither of these  approaches  were  possible in the analysis of the geopolitical nuclear deterrence stability between states,
because the behaviour of a state is much more complex with respect to the examples mentioned before
and, luckily,
no statistic of that kind of events can be made.\\
In this paper we are going to propose a physical model that can represent a simple deterrence equilibrium situation,  it is on based theory of unitary rational actors \cite{deterrence}.\\
This theory takes in account that a state is composed of a large number of people and a
detailed study of the dynamics of any individual it is impossible:
no equation can describe the behaviour of a single person and, also if this is possible,
the problem may have a too large number of variables to be solved. For this reason the theory of unitary rational actor was developed. But for the same reason this approach is  criticized \cite{no-unitary1} \cite{no-unitary2}.
In fact it is very hard to consider the whole range of psychological and sociologistic
factors, such as mental quirks, etc...\\
This situation, in Physics, is similar for gases, which are composed by a huge number of molecules and the study dynamics for one is impossible.\\
Then a common approach, in this case, is the thermodynamics one. We  only look to the macroscopic properties of the gas such as pressure, temperature and volume.\\
We want to emphasize that these quantities are strictly related to average behaviour of particles, but to estimate them, it  is not necessary to known the state of every body
that compose the gas.\\
Then in this sense thermodynamics could help to use theory of unitary rational actors  also considering some non-rational factors.\\
Firstly we are going to see how this model represents the Mutual assured destruction (MAD) theory \cite{mad1}\cite{mad2}\cite{mad3} if the influence of internal (economical, political,due to non state actors) 
instability and o conventional military forces
are not taken in account.
 Secondly we are going to consider the factors just mentioned in our model and we are going to study their influence on deterrence stability: MAD will not be anymore effective.\\ 
Finally we emphasize that the proposed model can simulate the reaction to the interest of a neighbour state or the international community on the deterrence equilibrium situation.  
This will be made through the concept of environment, in particular
this work we are going to suppose that is a thermal bath that keeps
the temperature of the system fixed, and temperature as the ``average`` interest of states outside the system on the system itself.\\
In particular we are going to apply this model to the India-Pakistan deterrence equilibrium and see how the influence of the soft international diplomacy can change the situation. 
\section{The model}
A gas can be described using a small number of macroscopic variables:
temperature ($T$) ,  pressure ($p$), numbers of moles  ($n$), volume occupied by a gas ($V$), ect.\\
The most important is the pressure, that represents the total pressure that a state could make on another state, this quantity can be composed 
of different contribution, e.g. an economical pressure due to an embargo or military pressure due to its nuclear arsenal, etc.\\
In order to study  equilibrium  we propose the following simple model
: consider  a cylinder, with volume $V$, divided in two parts ($V_1$ and $V_2$) by an adiabatic 
(no heat can be exchanged through the wall), free to move and frictionless wall.
Now suppose to fill the two volumes with two different gases and to fix the external temperature $T$, then at equilibrium the pressures of two gas must be the same:
 \begin{equation}
  p_1=p_2.
  \label{equilibrium}
 \end{equation}

 This is quite good in our context, because an equilibrium can exist if and only if the geopolitical pressure is the same between two different states. This is the reason why the pressure, 
 in our model, is be the discriminant on the existence of an equilibrium: if a state makes to much ( or too littll) 
pressure it would be very hard to reach equilibrium with another state.\\
To determine the pressure that a state is able to do, we use a state equation: a constitutive mathematical relation, thats gives the possibility to estimate one of the parameters if the others are knowns .\\
 Filling the two volumes with real gases (i.e. Van der Waals gases \cite{gas}), their pressure is:
\begin{equation}
 p_i=\frac{n_j R T}{V_j-n_j b_j}-\frac{a_j  n_j^2}{V_j^2},
 \label{vdw}
\end{equation}
\begin{centering}
 $$ j=1,2.$$
\end{centering}
Before looking deeply in equation \eqref{vdw} it is necessary to understand how we associate the parameters that describe a physical gas to a geopolitical context.

\begin{itemize}
 \item $R$ is a physical constant, that represents the nuclear deterrence of a state.\\
 \item $n_j$ is number of moles, it describe the amount of particles in the gas. In the purposed model it represents the capability of a state, an high value of $n$ indicates a
powerful (e.g. great gross national product) state.
 
 \item $T$ is the temperature of the gas and also of the external environment. Physically it describe the average agitation of particles that compose a gas.
 Pressure is directly proportional to temperature, so a state with high average agitation has more pressure.\\

 \item $V_j$ is the volume occupied by a gas. We will call it strategic volume. Fixed all the other quantities the volume of a ''geopolitical'' gas will represents its relative weight in an equilibrium situation.
 e.g. let's suppose that the volume of the cylinder, at equilibrium, is shared in such way that $\frac{V_1}{V}=0.666$ this means that for the equilibrium the state $1$ is twice more important that state $2$.
 We want emphasize that the relation $V_1>V_2$ does not means that state $1$ is more powerful than state $2$, but that its role is more crucial for the strategic deterrence equilibrium.

 \item $b_j$ corresponds to the intrinsic volume of atoms and molecules of the gas. This term keeps in account low range repulsive forces which oppose to the compression of the gas.\\
 In Geopolitical analogy it is  a critical parameter, because a gas can be compressed until it reaches
 $n_jb_j$ value, if this occurs it describes a dramatical situation, since not equilibrium can anymore occur. $b_j$ can represent the {\bf  internal instability} of a state. \\
 States with high value of $b_j$ are brought more easily to
 committing unrepairable mistakes, since a lower external pressure can bring this gas near $n_jb_j$ critical value.

\item $a_j$  represents a low range attractive force that reduce the pressure of a gas,  $a_j$ is positive definite.
In Geopolitics the situation is quite different; we assume that $a_j$ can be both positive or negative. \\
 If $a_j>0$ reduces the pressure of a gas, so positive value of $a_j$ can be interpreted
 as a {\bf  stability factor}. \\
Otherwise if $a_j<0$ is a {\bf  destabilizing factor}.

 \end{itemize}

  Let's look at equation \eqref{vdw} it is trivial that the pressure is composed by two terms.\\  
  First term gives high contribution to the geopolitical pressure if we are dealing with a large amount of strategic deterrence weapons (large n) or if the temperature is high.
  In geopolitical analogy we relate this contribution to the pressure made by the strategic nuclear deterrence of a state, 
In this case the therm gives a lower contribution if $b_j$ has a big value. So we can imagine that a state with internal problems is less influent on the balance of deterrence through its  strategical weapons.\\
  Then can be useful to parametrize $b_j$, as it is possible to distinguish from different contributions to the internal instability of state:
 \begin{equation}
  b_j= b_{eco}+ b_{nsa}+b_{stab},
 \end{equation}
every term of the last equation is positive, $ b_{eco}$ represents the contribution from {\bf  economical }problems, $b_{nsa}$ represents the contribution due to {\bf  non-state actors}
 and $ b_{stab}$ is the contribution due to {\bf  internal political} problems.\\
  The behavior of the second term,  $-\frac{a_in_i^2}{V_i^2}$,
  depends on the signs of the parameter $a_j$, as  if  $a_j<0$, it gives a positive contribute to the pressure, otherwise if $a_i>0$.\\
   Hence, as before, let's parametrize $a_j$:
 \begin{equation}
  a_j=a_{mil}+a_{dip},
 \end{equation}
where $a_{mil}$ is lesser than zero and represents the {\bf conventional military} contribution, while $a_{dip}$ is bigger than zero and represents {\bf diplomatic actions } that
reduce the pressure.
\section{Equilibrium equations}
  We solved  equation \eqref{vdw} respect $V_1$ in different limits to simulate different strategic behaviors.\\
  Firstly we impose $a_1=a_2=b_1=b_2=0$ then equilibrium condition becomes:
  \begin{equation}
   \frac{R_1n_1}{R_2n_2}=\frac{V_2}{V_1},
  \end{equation}
the last equation represents a  {\bf  strategic equilibrium equation}, since in this case we suppose that only the strategic nuclear weapons influence the equilibrium.
Last equation tells us that the ratio between strategic
volumes and number of strategic deterrence mechanisms multiplied by the capability, is the same.
The most important conclusion that can be draw is that equilibrium does always occur. Indeed the unique case 
in which a nuclear war can break out is the one where one of the two state does not have a nuclear strategic weapon. This is in good approximation with MAD theory.\\
The other case that we take in account is the one in which  $n_1=n_2=1$, $a_1=0$, $R_1=R_2$ and $b_2=0$.
 Equation  becomes \eqref{vdw}:
\begin{equation}
 \frac{n_1 R T}{V_1-n_1 b_1}=\frac{n_2 R T}{V_2-n_2 b_2}-\frac{a_2  n_2^2}{V_2^2}.
 \label{vdwIndiaPak}
\end{equation}
This equation represents a {\bf  simplified deterrence and strategic equilibrium} of an asymmetric Geopolitical situation.\\
The hypothesis was made to fit a existing situation of  equilibrium among India and Pakistan.
For sake of relevancy to the real situation, from now we interchange index $1$ with index $p$ (since refers to Pakistan) and index $2$ with $i$ (refers to India).

\begin{figure} 
\centering
\includegraphics[width=13cm,natwidth=610,natheight=642 ]{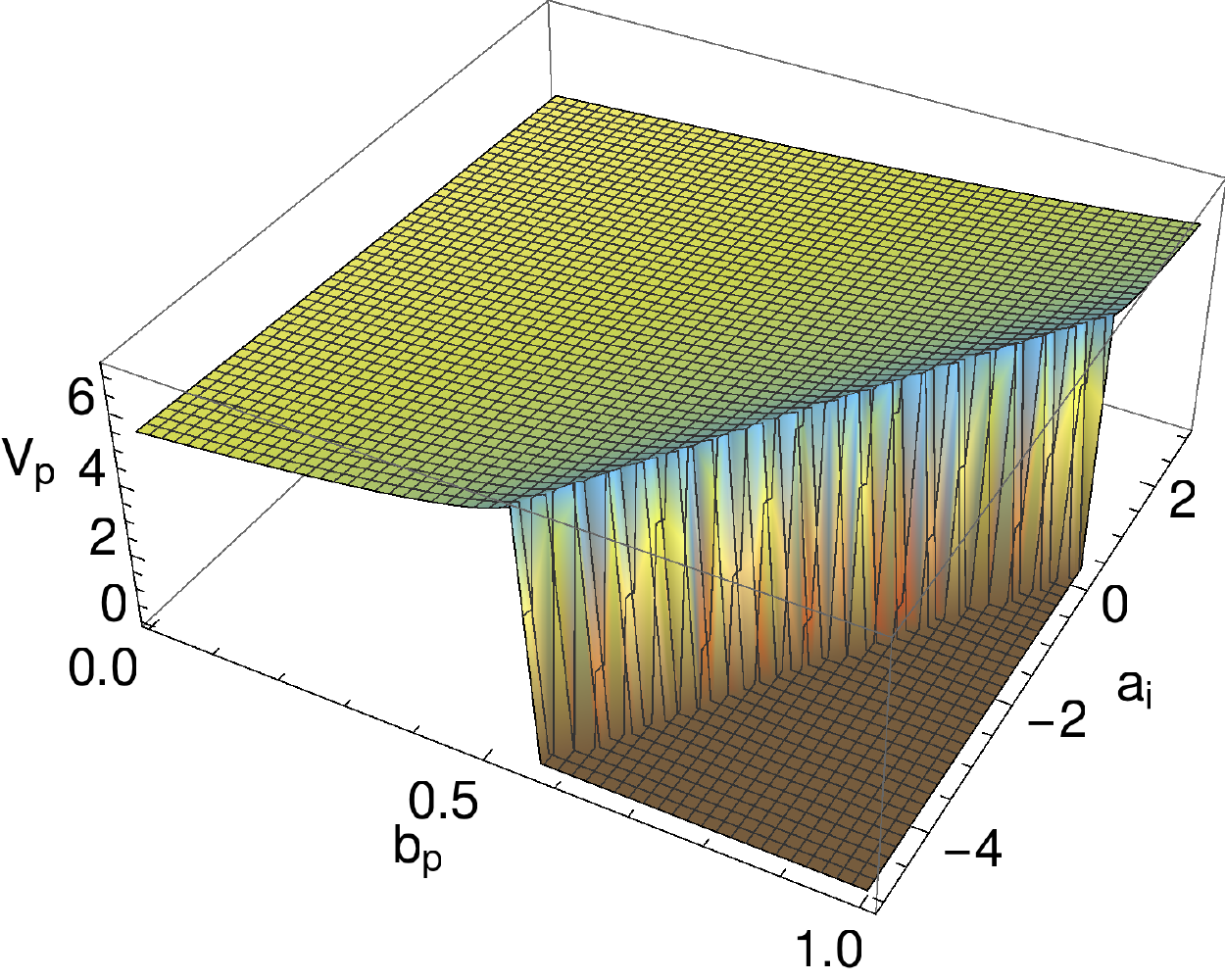}
\caption{$V_p$ in function of $a_i$ and $ b_p$ with $T=1 K$ and total volume $V=10 L$. }
\label{3d}
\end{figure}

\begin{figure}
\centering
\subfloat[][\emph{$V_p$ in function of $b_p$ with $a_i=0.5$, $T=1 K$ and total volume $V=10 L$.}.]
{\includegraphics[width=.45\textwidth]{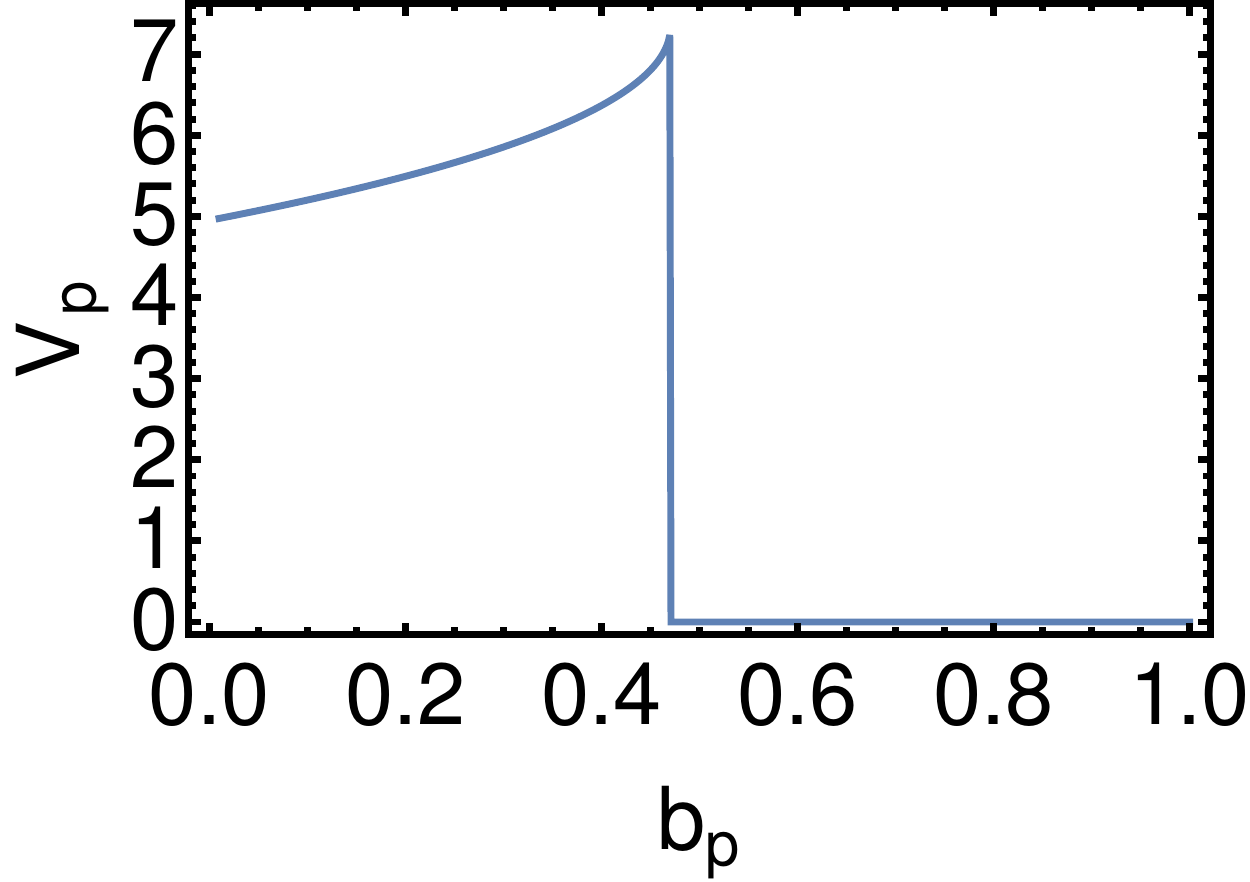}} \quad
\subfloat[][\emph{$V_p$ in function of $a_i$ with $b_p=0.5$, $T=1 K$ and total volume $V=10 L$.}]
{\includegraphics[width=.45\textwidth]{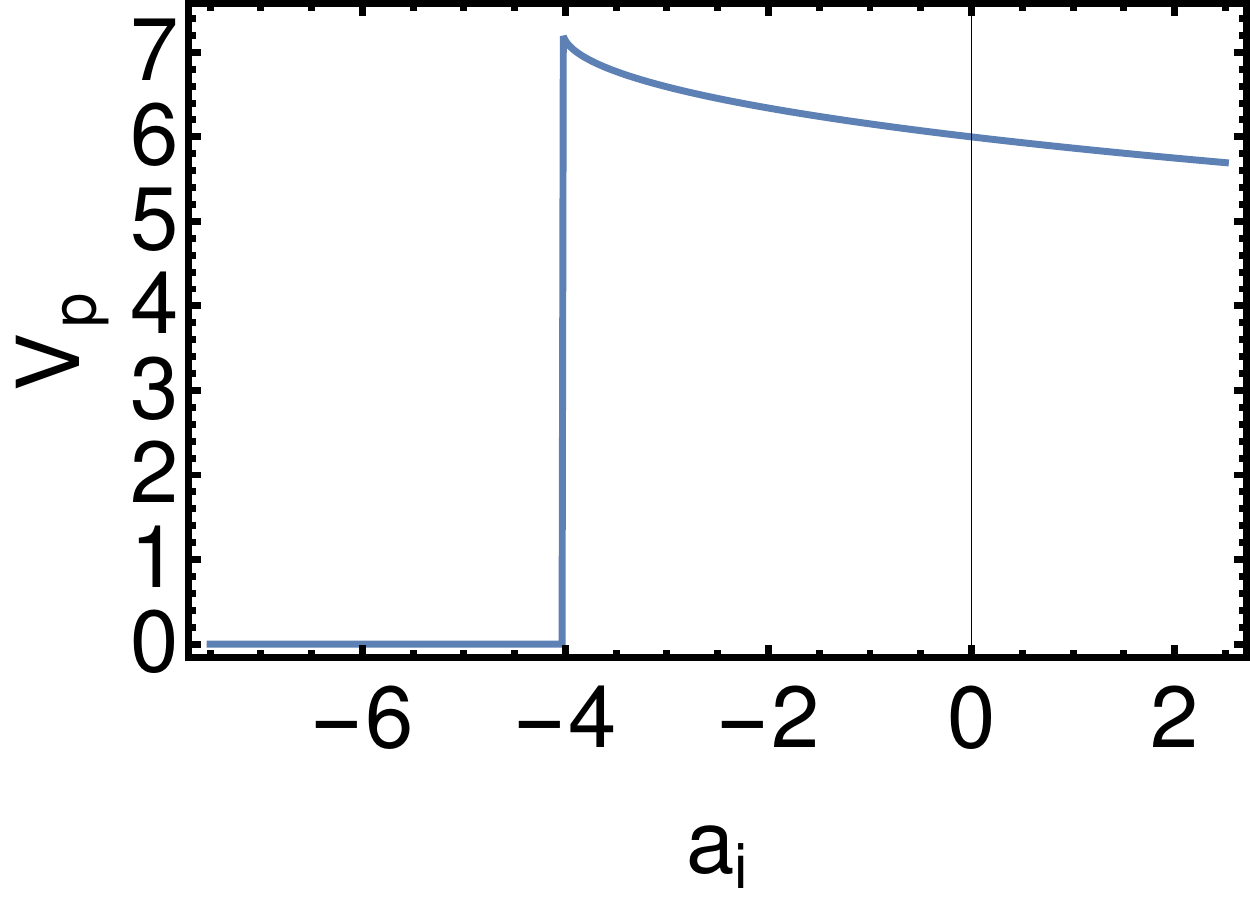}} \\
\caption{Sections of figure \ref{3d}. }
\label{vs}
\end{figure}

Then imposing  $R_1=R_2$ and $n_i=n_p=1$ we assume that both states have enough 
nuclear deterrence to annihilate each other and the state capability it is similar (not true in reality but the first aim of this work it is tu study the influence of others factors on equilibrium, the study 
of assymetric nuclear and economical situation will be adress to fututre works).
Furthermore $a_p=0$ means there is not a substantial conventional military pressure  on India made by Pakistan.\\
Finally we set $b_i=0$, India in this model does not have a critical volume.
This means  on one hand that any pressure from Pakistan can not convince India to make the first use of a nuclear weapons, on the other hand that India does not have rouge state behaviors
or internal economical  problems of concern \cite{india-pak}.\\
We solve this equation with respect to $V_p$, fixing $T=1K$, $V=10L$, using as parameters, representing the Geopolitical situation, $a_i$ and $b_p$. 
In figure \ref{3d} it is possible to see one solution (system has two roots, one for our purpose is meaningless).\\
In figure \ref{vs}, we can see sections of the previous graph, obtained fixing one parameter.\\
Looking at these graphs one should keep in mind that $V_p=0$ means that in such points of parameters space is not possible to find Physical (e.g. real) solutions of equation (\ref{equilibrium})
(we set $V_p=0$ in these cases in order to make graph, a priori this situation is the liquefaction of one of the two gases and volume can be a complex number).\\
Then Geopolitically in these points no kind of strategic and deterrence equilibrium can be achieved.\\
To use this graph one must estimate $a_i$ and $b_p$ of the real situation, then using figure \ref{3d} it is possible to say if with these estimated values there exists
a strategic and deterrence equilibrium and how far is  this point from  the cliff.\\
For example, one looking at the real situation estimates  $b_p=0.5$, then figure \ref{3d} shows for which value of 
$a_i$  there is an equilibrium. \\
In particular let's consider figure \ref{vs}.b where $b_p$ is fixed at $0.5$,  then it is possible says for $a_i\leq -4$ India's
conventional military pressure is too much for the existence of an equilibrium. Otherwise if $-4<a_i<0$ we have a conventional military pressure from India,
by the way equilibrium can occur. 
It is important to notice that an $a_i$ decreasing represents the escalation dominance of India, an increase of conventional military power in order to assure a  rapid dominance in war scenery.
From these results we can say that escalation dominance could be
a threat for equilibrium.\\
Eventually for $a_i>0 $,
India tries a diplomatic d\'etente, for $b_p=0.5$ there is always equilibrium.\\
Another example is figure \ref{vs}.a, in this case suppose,  a priori, to estimate $a_i=0.5$, then only if $b_p<0.75$ an equilibrium is possible.\\
What further information can be extracted from figure \ref{3d}?\\
Since $a_i$ can be both positive and negative we can distinguish between two cases.\\
If $a_i > 0$ in this case the term $-\frac{a_i}{V_i^2}$ gives a negative contribution to $p_i$, then this region represents a situation where India tries a diplomatic d\'etente.
An interesting fact is that also in this case a nuclear war can occur when $b_p$ is large enough. This means that internal (economical or political) instability or 
the action of non-state actor are sufficient to cause a dramatic situation even thought India's distress actions.\\
If $a_i<0$ the term $-\frac{a_i}{V_i^2}$ gives a positive contribution to $p_i$. This behaviour simulates that the conventional military force of India makes some
pressure on Pakistan.  Luckily, also in this case equilibrium can occur, obviously bigger $a_i$ is lesser  the value of $b_p$ must be in order to  make equilibrium possible.\\
Then our model shows that the most important issue for reaching a geopolitical stability in South-Asia area is 
the presence of non-state actors or  internal instability of a state. The presence of nuclear powers in the area is not mainly cause of instability and it is not sufficient
 to be cause of a war. \\
 Note also that the strategic volume of Pakistan is bigger than India's one, this means that the role of Pakistan is more important in this situation and the priority should be the ``stabilization'' of
 Pakistan internal politic.\\
 These results are important since our model supposes that a state is a unitary actor but shows that factors due to instability or conventional military forces are fundamental to reach an equilibrium,
 usually this kind of factors  are taken away using the unitary actors theory.

\section{The role of temperature.}
\begin{figure}[htp] 
\centering
\includegraphics[width=14cm ]{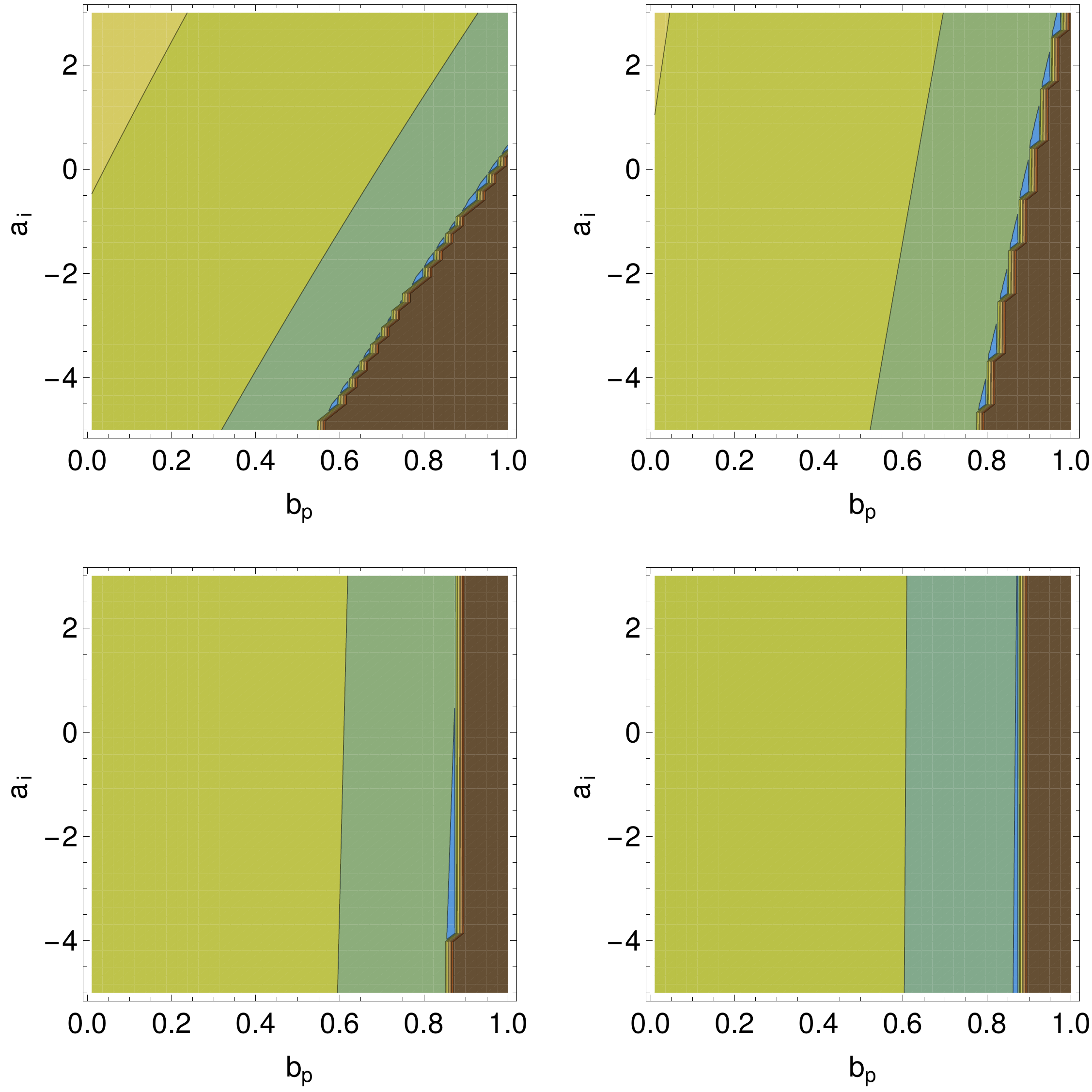}
\caption{$V_p$ in function of $a_i$, $b_p=$, at different temperature, seen from above.
At the first row starting from left we  $T=1$ [K] and $T=3.5$ [K], indeed the second row $T=25$ [K] and $T=100$ [K].
The brown area are places in parameters space where $V_p=0$.}
\label{temp}
\end{figure}

As it was said before one aim of this work is to study a geopolitical situation between two states considering also the interference made by other states.\\
The main advantage in using a thermodynamics toy model is the reservoir.
In fact if we put the system into a reservoir, then the temperature of the system becomes the same temperature of
environment, then we are considering a thermal bath.\\
So it is possible to simulate the action of neighbour state as a different temperature.
In previous section we set arbitrary $T=1K $, that condition  will represent a situation where the system (i.e. the two state)
has a low interaction with other states: the attention of international  establishment or of neighbour state is  elsewhere;
then we can figure out that a temperature increase represents a political interest on the system from external state actors,
this can be  a good approximation because before we said that $T$ is the average 
agitation of the components of a gas, then  attention of external states can enlarge the worries of the state that compose the system, so this will cause a temperature increase.\\
This analogy has advantages and disadvantages, actually it is possible to study a problem that considers more than two states without working with a multi-bodies (three or more) problem, but 
this assumption works if and only if we are dealing with external actors that are not heavily influenced by changes in the system. 
In other words reservoir does not changes its temperature if there are changes
in the system.
In Geopolitics this is a good approximation when we suppose that an external actor 
is very powerful (economically and military) in comparison of states in the system. \\
For these reasons in
 India-Pakistan equilibrium can be again a good example to test our model, because the role of reservoir can be made by Internatiol soft diplomacy,
infact we can say that the international establishment is not affected too much by the situation between India and Pakistan.
For sake of understanding we set arbitrary $T \in [1K,100K]$.\\
In figure \ref{temp} is possible to see the same figure \ref{3d}, viewed from above,  evaluated at  different temperatures (remember that figure \ref{3d} is evaluated at $T=1K$). \\
 Look in particular \ref{temp}(d)    with external high  temperature ($T=100K$) the model  shows a particular behavior.\\
 First of all, equilibrium 
can exist for any value of $a_i$, from a mathematical point of view, with this temperature, the contribution of the term $\frac{a_i}{V_i^2}$
is small compared to the others  (in  both cases, if it is an stabilizing factor or not).\\
This fact could mean that if the international establishment gets interested  in the area
(then $T$ is high), then India spreads its energy and loses the interest in making pressure on Pakistan with conventional military power.\\
Hence this intervention, in our model, reduces the pressure on Pakistan.\\
Note also that in some points in the plane $(a_i,b_p)$ where at $T=1K$  equilibrium solutions do not exists, for  high temperatures these points may have solution.
For these reasons  an international intervention, in this model,
is a stabilizing factor.\\
On the other hand, in this case  ($T=100K$), there exists a critical value of $b_p$ (i.e. $b_ {cri}=0.71$), when $b_p>b_{cri}$
equilibrium can not occur independently on any actions.\\
This represents that there exists a break point of Pakistani problems, if Pakistan has too many internal problems (economical or political) or too much rouge state behavior a 
potential war might breaks out, regardless of international peacemaking efforts.\\
In other words, a international strong intervention could dramatically change the scenery, it could make inefficient the effects of India's conventional military pressure on Pakistan
(i.e. the equilibrium point exists despite the value of $a_i$), for this reason 
due to this action it is possible to have some equilibrium points that would not  exist otherwise.
However Pakistani problems can always be source of instability for all South Asia area.\\ 

\section{Conclusion}
The proposed model through gases enables to simulate different behaviors.
In one limit, where no ``weight'' is associated to conventional military power or to internal instability source,  it simulates the MAD theory, since no war might break out (equilibrium always exists)  
until both state has a nuclear strategic deterrence.\\
Otherwise this model is able to represents an asymmetric situation where one state has a large conventional military power and no internal problem and the other has only internal stability issues and a negligible
conventional military
compared to the other.
In this case using this model to represent the deterrence equilibrium between Pakistan and India shows how this asymmetry is source of instability for the whole area.
The model is able to incorporate these factors even if it is based on unitary actors theory, in particular  we saw the effect of Pakistani non-actor state, India's diplomacy and
 escalation dominance. We conclued that the best option for India is the diplomatic one but the key role is adressed to Pakistan's internal problem.\\
 \\
The last advantage of this model is the possibility to simulate the action of a powerful neighbour state through the temperature of the thermal bath.
In this case we showed that international  soft diplomacy intervention could be a strong stabilizing factor for the Pakistan-India equilibrium issue, but the key role is again addressed to the Pakistani internal problems.
These problems are discriminant for the existence of the equilibrium and could be a source of war even with a strong external intervention, so  our idea is  that priority should be given to solve Pakistani
internal problems to reach a long term stability in the South-Asia area.

 \section*{Acknowledgments}
Useful discussions with Silvia Cassina are gratefully acknowledged.

\end{document}